\documentstyle[preprint,aps]{revtex}
\begin{document} 
\newcommand {\ybco} {\rm YBa_2Cu_3O_{6+x}}
\newcommand {\ybcos} {\rm YBa_2Cu_3O_7}
\newcommand {\lsco} {\rm La_{2-x}Sr_xCuO_4}
\newcommand {\bsco} {\rm Bi_2Sr_2CaCu_2O_8}
\newcommand {\bscoo} {\rm Bi_2Sr_2CuO_x}
\newcommand {\bscot} {\rm Bi_2Sr_2CaCu_2O_x}
\newcommand {\ybce} {\rm YBa_2Cu_4O_8}
\newcommand {\Tc} {T_c}
\newcommand {\beq} {\begin{equation}}
\newcommand {\eeq} {\end{equation}}

\title{Model of C-Axis Resistivity of High-$\Tc$ Cuprates
}
\author{Yuyao Zha,  S. L. Cooper and David Pines}
\address{
Department of Physics, University of Illinois at Urbana-Champaign,
1110 West Green Street, Urbana, IL 61801
}
\date{\today}
\maketitle

\begin{abstract}
We propose a simple model which accounts
for the major features and systematics of experiments on the $c$-axis
resistivity,
$\rho_c$, for $\lsco$, $\ybco$ and $\bsco $. We argue that the $c$-axis
 resistivity can be
separated into contributions from in-plane dephasing and the
$c$-axis ``barrier'' scattering processes, with the low temperature
semiconductor-like behavior of $\rho_c$
arising from the suppression of the in-plane density of states
measured by in-plane magnetic
Knight shift experiments. We report on predictions for $\rho_c$
in impurity-doped $\ybco$ materials.
\end{abstract}
Although there is currently no consensus\cite{models}
as to the important mechanisms
contributing to $c$-axis transport in high temperature superconductors,
recent transport and optical experiments of $\bscot$, $\ybco$, and $\lsco$
\cite{bsco,uchida2,uchida1} reveal a number of key features that must be
accounted for in any successful
model of the interlayer charge dynamics in the layered cuprates: First,
$\rho_c(T)$ in $\bsco$ and underdoped $\lsco$ and $\ybco$ have a
semiconductor-like temperature dependence $(d\rho_c /dT <0)$ at low
temperatures, and a linear-in-T dependence at high temperatures. The
crossover temperature, $T^*$, between these two regimes decreases with
increased doping in $\ybco$ and $\lsco$. Importantly, $c$-axis optical
measurements\cite{timusk} show that the semiconductor-like resistivity
``upturn'' in
underdoped $\ybco$ is actually associated with a uniform suppression of
the optical conductivity below $\sim 300 cm^{-1}$. These data suggest that
the $c$-axis conductivity scales at low frequency with the Knight shift,
which is proportional to the in-plane density of states: $K_s \propto N(0)$.
Second, both $\lsco$ and $\ybco$ exhibit a strongly
doping-dependent mass anisotropy, and a possible crossover from quasi-2D
to 3D transport behavior at high doping, that arises in part from
doping-induced structural changes\cite{cooper2,cooper}.

In this paper, we present a simple phenomenological model of the $c$-axis
resistivity in the layered cuprates that describes the key elements of
interlayer transport in the cuprates with reasonable parameters. Before
examining the  mechanisms contributing to $c$-axis transport
in the cuprates, it is important to point out that experiment evidence
suggests that, except perhaps for the overdoped cuprates, $c$-axis transport
in high $\Tc$ superconductors is incoherent.  For example, typical estimates
of the $c$-axis scattering rate in $\ybco$ give $1/\tau_c > 1000 cm^{-1}
$\cite{cooper2}, while
optical measurements indicate a $c$-axis plasma frequency of
$\omega_{p\perp} \sim 40 meV$ in
fully oxygenated $\ybcos$, and $\omega_{p\perp} < 10 meV$ in underdoped
$\ybco$\cite{cooper}.  These values
suggest that the $c$-axis mean free path is of order or less than the $c$-axis
lattice spacing, i.e., $c$-axis conductivity in the cuprates is below the
Ioffe-Regel limit, and hence $c$-axis transport is incoherent.

        One important contribution to $c$-axis transport in the cuprates
is expected
to arise from electron scattering in the ``barrier'' layer between
${\rm CuO_2}$ ``cells'' (i.e., layers, bilayers, trilayers, etc.).
For example, $c$-axis
Raman scattering measurements provide evidence that carriers hopping
between layers scatter from $c$-axis optical phonons associated with the
barrier in $\ybco$\cite{cooper2}, while Littlewood and Varma have pointed
out the
likelihood that static impurities in the barrier layer provide an important
source of scattering for $c$-axis transport in the cuprates\cite{models}.
 A phenomenological expression for this contribution to $c$-axis transport can
be written as:
\beq
        \sigma_c^{(1)} = N(0)\frac{e^2d^2}{\hbar^2} t_{\perp}^2 \tau_c
\label{incoh}
\eeq
where $d$ is the interlayer spacing, N(0) is the in-plane density of states,
$t_{\perp}$ is the interlayer coupling, and $\tau_c$ is the $c$-axis scattering
time.

        On the other hand, $c$-axis transport measurements of $\lsco$ and
$\bscoo$
yield $\rho_c \propto \rho_{ab}$ at high temperatures\cite{uchida1,hou},
suggesting that
scattering or
fluctuations in the planes may dominate $c$-axis transport in this regime.
This
contribution to $c$-axis transport can be written\cite{legget},
\beq
\sigma_c^{(2)}=N(0)\frac{e^2d^2}{\hbar^2}t_1^2\tau_{ab}
\label{ab}
\eeq
where $\tau_{ab}$ can be derived from the planar conductivity,
\beq
\sigma_{ab}=\frac{\omega_{p\parallel}^2}{4\pi} \tau_{ab}
\label{tauab}
\eeq
and the temperature-independent quantity, $t_1$, measures the effectiveness of
 planar scattering processes to $c$-axis transport.

Because Eqs.(\ref{incoh}) and (\ref{ab}) describe independent physical
processes, it seems natural to consider the corresponding tunneling
and/or
scattering mechanisms as additive in the resistivity. We are thus led to
propose the following
expression for $\rho_c$,
\beq
\rho_c=\frac{\hbar ^2}{N(0)e^2d^2} \big( \frac{1}{t_1^2 \tau_{ab}}+\frac{1}
{t_{\perp}^2\tau_c} \big).
\label{model}
\eeq
In the limit that one mechanism or the other is dominant, Eq.(\ref{model})
yields the
corresponding conductivity given in Eqs.(\ref{incoh}) and (\ref{ab}). We now
proceed to use Eq.(\ref{model}) to analyze $c$-axis transport experiments.

We consider first the $\lsco$ system. We obtain the temperature-dependent
density of states $N(0)$, from the recent analysis by Barzykin,
Pines and Thelen\cite{victor}, who extract the
temperature-dependent uniform
susceptibility, $\chi_o(T)$, from Knight shift measurements and scaling
arguments on this system;
in deriving $N(0)$, we neglect Fermi liquid corrections, determining $N(0)$
from
$\chi_o(T)=-\mu_B^2 N(0)$ where $\mu_B$ is the Bohr magneton. The density of
states $N(0)$ obtained from the
Knight shift data of Ohsugi $et~al$\cite{kslsco} is shown in Fig. 1.
We determine $\tau_{ab}$ by using Eq.(\ref{tauab})
and optical measurements of $\omega_{p\parallel}$\cite{cooper}. An
independent measure of $t_{\perp}$, the
interlayer hopping amplitude, can be obtained from the $c$-axis plasma
frequency, $\omega_{p\perp}$, measured in optical or penetration depth
experiments\cite{cooper}.
This leaves us
with a two-parameter fit to the data for each hole doping, these parameters
being $t_1$ and $1/\tau_c$. Our results
are given in Fig. 2 and Table I, and we comment on them briefly.

We first
note that at each doping level, the high temperature behavior of $\rho_c$
is determined entirely by the planar conductivity, according to Eq.(\ref{ab}),
so that $t_1$ is completely determined by $\rho_{ab}$. The values of $t_1$
obtained in this way are independent of hole
concentration, within 10\%.
Secondly, both $t_{\perp}$ and $\tau_c$ are independent of temperature, and
the barrier layer scattering contribution described by Eq.(\ref{incoh})
dominates at
sufficiently low temperatures.
As might have been expected, $\tau_c$ is essentially independent of doping,
while $t_\perp$, which increases with increasing hole concentration, displays
a strong dependence on hole doping where $t_{\perp}
\sim x^{\alpha}$ with $\alpha > 2$. According to
Eq.(\ref{model}), the crossover temperature $T^*$,
which separates the $d\rho_c/dT <0$ and $d\rho_c/dT>0$ regimes, occurs
when the two terms in the parenthesis of Eq.(\ref{model}) are equal.
After some simple algebra, we find that at $T^*$,
we have
\beq
\frac{\rho_c(T^*)}{\rho_{ab}(T^*)}=\frac{\hbar ^2}{N_{T^*}(0)e^2d^2}
\big( \frac{2}{t_1^2 \tau_{ab}}\big)\Big/\big(\frac{4\pi}
{\omega_{p \parallel}^2\tau_{ab}}\big)
=\frac{\hbar^2\omega_{p \parallel}^2}{2\pi N_{T^*}(0)e^2d^2t_1^2 }
\label{star}
\eeq
where $N_{T^*}(0)$ is the density of states at T=$T^*$. Roughly, both
$N_{T^*}(0)$ and $\omega_{p \parallel}$ increase
with doping concentration, while all the other quantities on the right
hand side of
Eq.(\ref{star}) are doping independent. It
turns out that the anisotropy ratio $\rho_c/\rho_{ab}$ at the crossover
temperature $T^*$  is nearly independent of
hole doping
for each individual cuprate. Thus for $\lsco$, Nakamura
$et~al$\cite{uchida1} found $\rho_c/\rho_{ab}
\sim 300$ at T=$T^*$, for doping levels of $x=0.10$, $x=0.12$, $x=0
.15$, and
$x=0.20$.

We now apply the same method of data analysis to $\ybco$ experiments.
Our results are shown
in Figs. 3 and 4, and  Table I. In Fig. 3, we give our results for $N(0)$,
 obtained from the analysis by Ref.\cite{pines} of the
Knight shift experiments of Ref. \cite{ksybco}.
The nearly doping independent results for $\tau_c$ shown in Table I
are taken from fits
to optical and far infra-red experiments\cite{cooper2}. Again we see that
$t_1$ is nearly
independent of hole concentration, while $t_{\perp}$ increases with
hole doping as $t_{\perp}
\sim x^{\alpha}$ where $\alpha > 2$. The anisotropy ratio
at the crossover temperature $T^*$ remains almost doping independent.
According to
the data of Ref. \cite{uchida2}, this ratio is $\rho_c/\rho_{ab} \sim 100$
at T=$T^*$ for
different doping levels.

A further test of our model is provided by the $c$-axis resistivity
measurements on $\bsco$ shown in Fig. 5, which display
the same qualitative dependence on temperature. Our theoretical results are
obtained by deducing $N(0)$ from the Knight shift experiments of Walstedt
$et ~al$\cite{walstedt}, $\tau_{ab}$ from the measurements by Martin
$et~al$ of $\rho_{ab}$ and $t_{\perp}$ from Ref.\cite{cooper}. The agreement
between our model calculation and
experiment is again seen to be satisfactory. The parameters used in
making the fit given in Table I.

It is instructive to compare the variations in $t_1$ and $t_{\perp}$ on going
from one system to another. The fact that $t_1$ is the largest
in $\ybco$, somewhat smaller in $\lsco$ and smallest in $\bsco$ suggests
that although independent of doping, the dephasing of the $ab$-plane enters
$\rho_c$ in a way which is related to the interlayer (unit cell) distance in
these
cuprates. Another observation from Table I is  the systematic
behavior of the $c$-axis
``barrier'' scattering rate $1/\tau_c$: the scattering is strongest in
$\ybco$, which contains
two apical oxygens per unit cell; weaker
in $\lsco$, which has one apical oxygen per unit cell; and significantly
reduced in $\bsco$,
which does not contain any apical oxygens. Our result for $1/\tau_c$ is
also consistent with the Drude fitted results from optical and
Raman experiments for $\ybco$\cite{cooper2,cooper}. All these facts support
the
suggestion that the $c$-axis barrier scattering
 originates most likely from the $c$-axis apical oxygen phonons.

$c$-axis transport measurements on impurity-doped systems provide a direct
test of our description of $\rho_c$ in Eq.(\ref{model}). We consider two types
of impurities: Zn and Co, both of which have been doped into $\ybcos$. Zn is
known to go in as a planar substitute for Cu.
To first order, it does not modify the planar
density of states $N(0)$, or the
in-plane scattering rate $1/\tau_{ab}$, nor will it influence $t_{\perp}$
and $\tau_c$. Therefore, we predict that for Zn doped $\ybcos$,
the changes which take place in the $c$-axis transport will mirror the
comparatively small increase, $\delta \rho_{ab} \propto n_{\rm Zn}$, found
for this system\cite{uchida3}.
On the other hand, when Co is doped into $\ybcos$ up to a 2.5\% doping level,
it does not influence $\Tc$\cite{co}, and may plausibly be assumed to
substitute for chain Cu atoms. It will therefore not affect $t_1$,
$\tau_{ab}$, or $N(0)$,
and we may expect $\sigma_c^{(2)}$, Eq.(\ref{ab}) to be unaffected by Co
substitutes to this level. Nevertheless, we would expect that
$t_{\perp}$,
which is sensitive to the chains, will be reduced, thus increases the
magnitude of the barrier scattering contribution to $\rho_c$,
and we expect $\rho_c$
to have a temperature dependence somewhat  similar to that of $\ybcos$:
linear-in-T
for the whole temperature range, with the same slope
but a larger residual resistivity at T=0.

In conclusion, we have presented a model for $c$-axis
resistivity of high $\Tc$ cuprates,
in which both in-plane dephasing and the $c$-axis ``barrier'' scattering
contribute
to the $c$-axis resistivity. The behavior of the ``barrier'' scattering is
consistent with that of the $c$-axis apical oxygen phonons.
Our model fits quite well with
the existing
data on $\lsco$, $\ybco$ and $\bsco$.

We acknowledge stimulating conversations with A. J. Leggett, Q. M. Si,
A. G. Rojo, D. M . Ginsberg, K. Levin, J. T. Kim and  L. Y. Shieh.
This work is supported by NSF-DMR-91-20000.

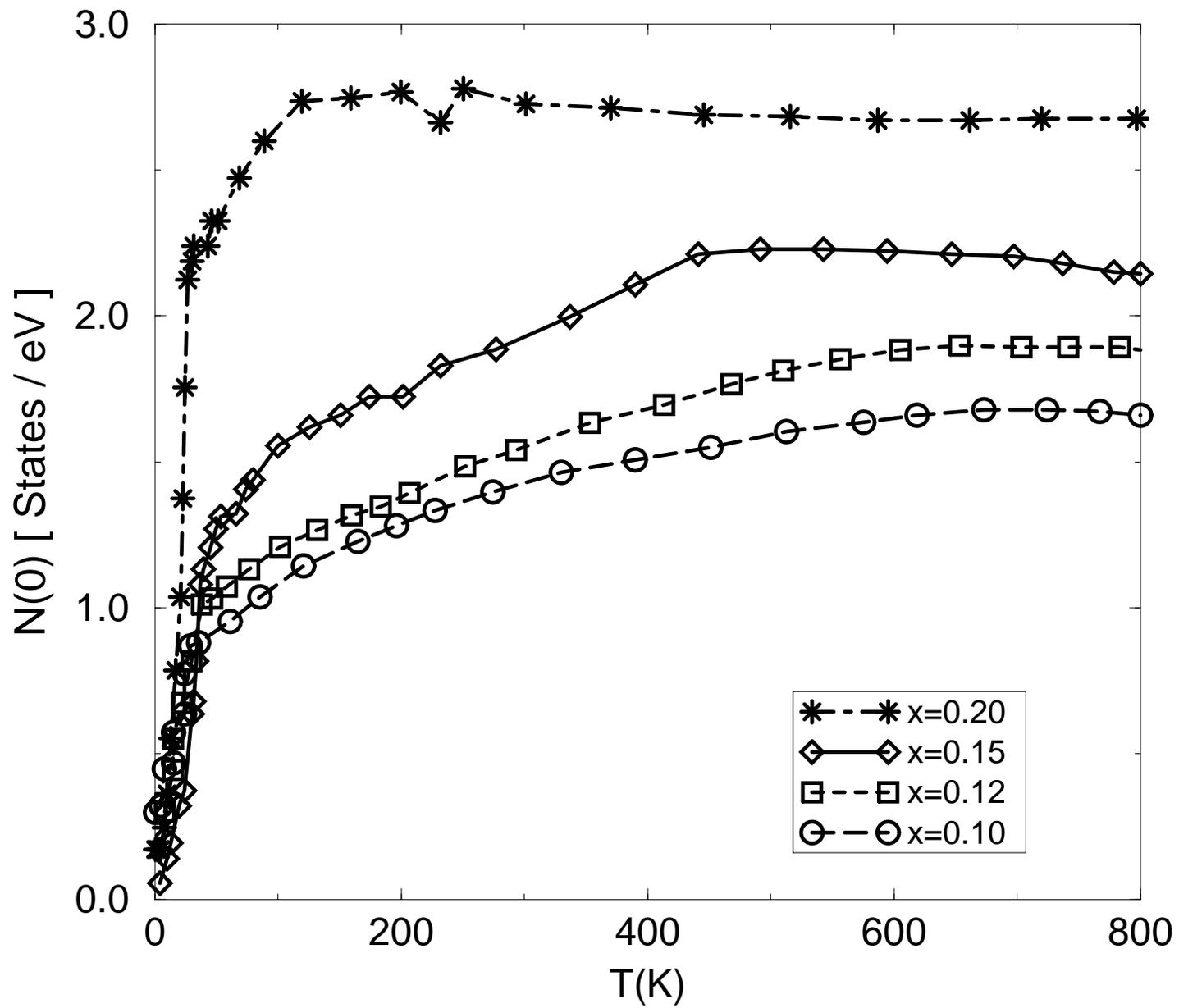
\begin{figure}
\caption{
The planar density of states of $\lsco$ obtained from the Knight shifts
data of Ohsugi  $et~al$
\protect\cite{kslsco}, following the scaling analysis of
Ref. \protect\cite{victor} .
}
\end{figure}
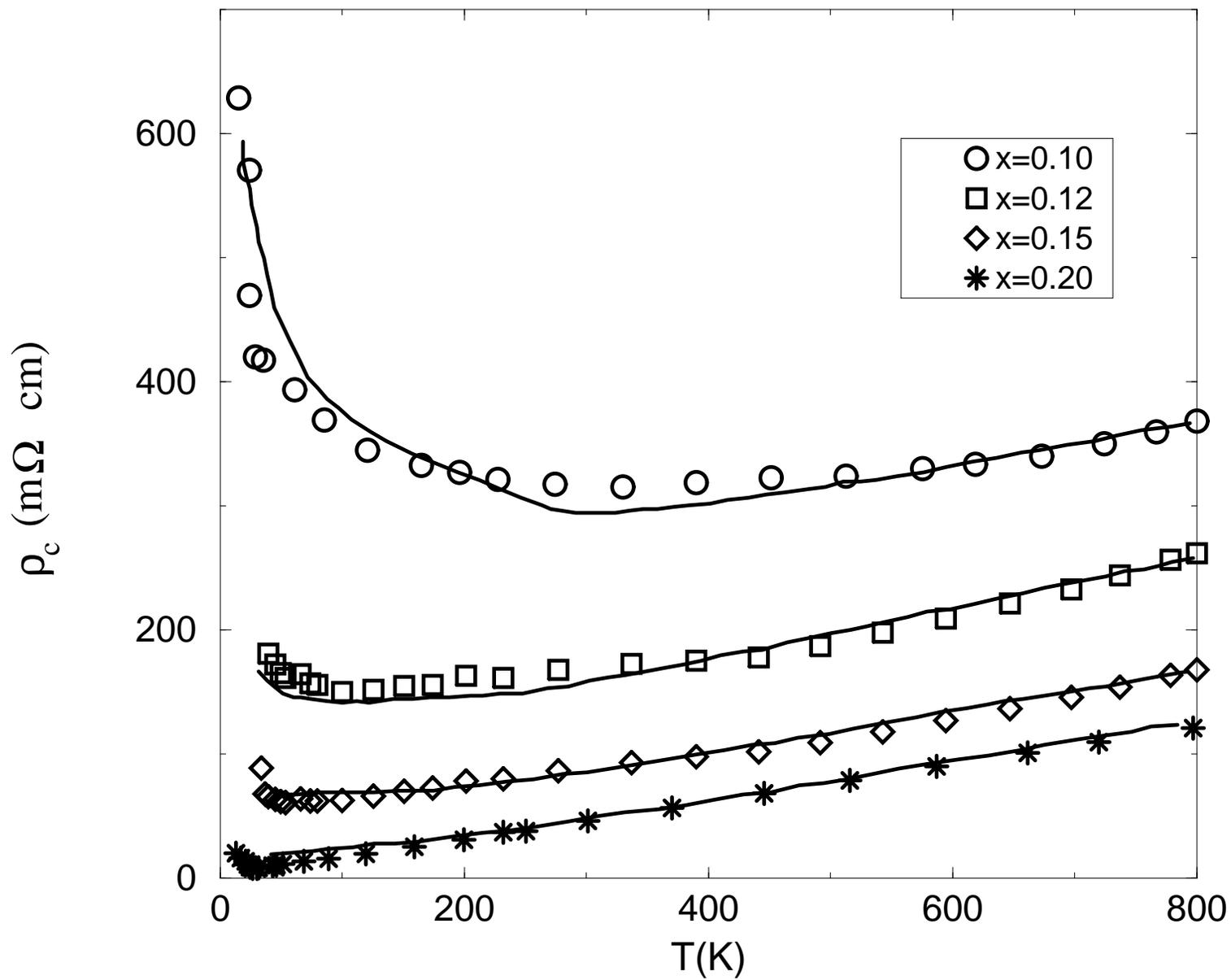
\begin{figure}
\caption{
Calculated $\rho_c$ for $\lsco$ at different doping levels (symbols), plotted
against the experimental data of Nakamura $et~al$\protect\cite{uchida1}
(solid lines).
}
\end{figure}
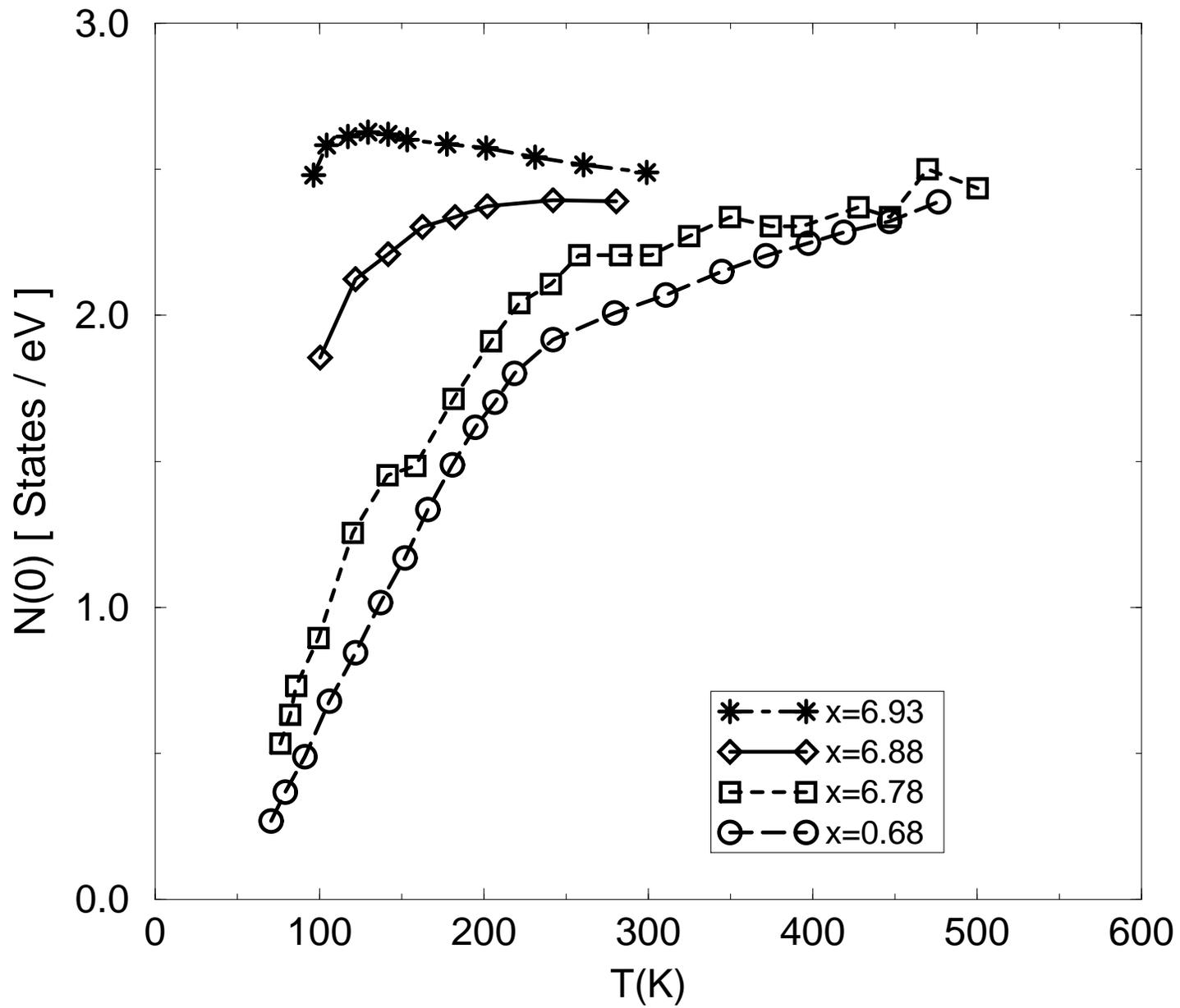
\begin{figure}
\caption{
The planar density of states obtained from the analysis of Ref.
\protect\cite{pines} for $\ybco$,
here we use $N(0)$ of ${\rm Y_{0.9}Pr_{0.1}Ba_2Cu_3O_7}$ for
${\rm YBa_2Cu_3O_{6.88}}$. The $N(0)$ for ${\rm YBa_2Cu_3O_{6.68}}$ is an
estimate from the scaling arguments given in Ref. \protect\cite{pines}.
}
\end{figure}
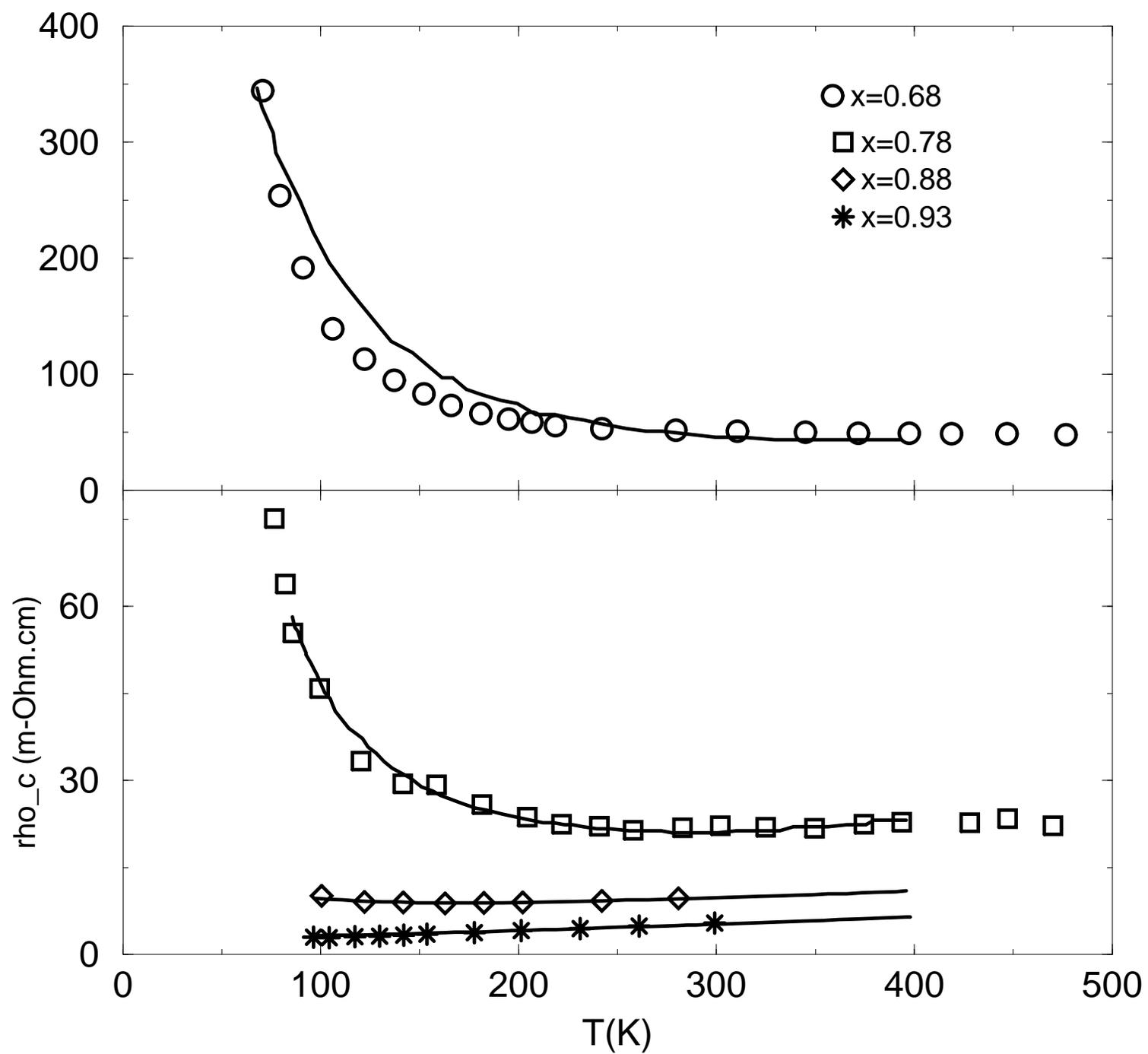
\begin{figure}
\caption{
Calculated $\rho_c$ for $\ybco$ at different doping levels (symbols), plotted
against the experimental data of Takenaka $et~al$\protect\cite{uchida2}
(solid lines).
}
\end{figure}
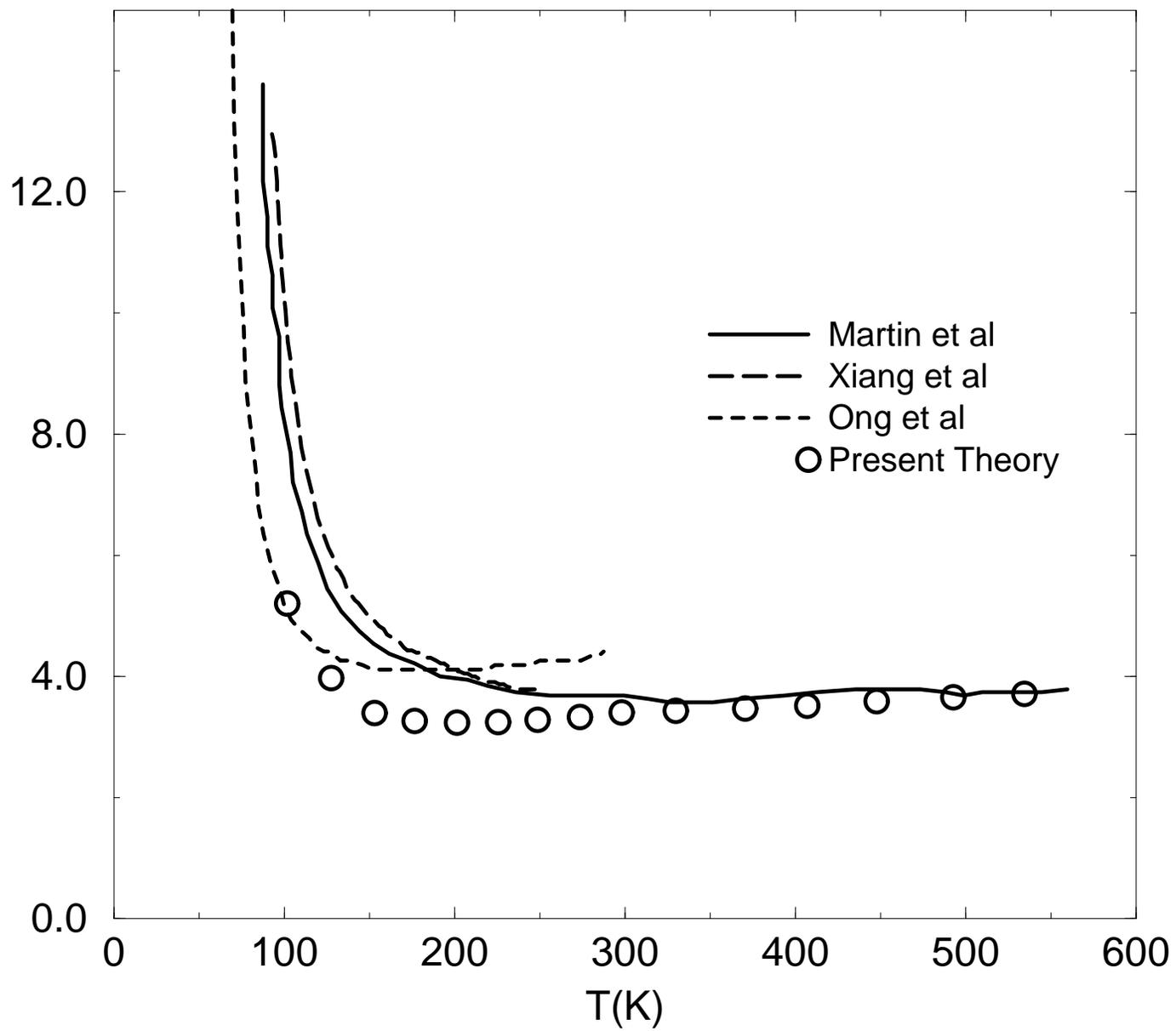
\begin{figure}
\caption{
Calculated $\rho_c$ for $\bsco$ (circles) in comparison with experimental data
by different groups\protect\cite{bsco}. $N(0)$ is obtained from the
scaling analysis of the Knight shift measurement of
Ref. \protect\cite{walstedt}.
}
\end{figure}

\begin{table}
\caption{Relevant Numbers}
\begin{tabular}{|c||c|c|c|c|c|c|c|}
&$\omega_{p\parallel}$&$t_1$ & $1/\tau_{ab}$ & $t_\perp$ & $ 1/\tau_c$
& $(t_1^2 \tau_{ab})^{-1}$ [at 300K]& $ (t_{\perp}^2 \tau_c)^{-1}$\\
& [eV]& [meV] &[meV]& [meV] & [meV] & [meV$^{-1}$]  &[meV$^{-1}$]
\\ \hline
${\rm YBa_2CuO_{6.68}}$&0.8& 14.4&$\sim 2kT$ & 7.3& 100 & 0.25 &1.88\\
${\rm YBa_2CuO_{6.78}}$&1.0& 14.1&$\sim 2kT$ &10.4& 100 &0.26&0.92\\
${\rm YBa_2CuO_{6.88}}$&1.18&  18 & $\sim 2kT$ & 16.7 & 100&0.16&0.36 \\
${\rm YBa_2CuO_{6.93}}$&1.4& 16.6&$\sim 2kT$ &30-40& 109 &0.19&0.089\\
\hline
${\rm La_{1.90}Sr_{0.10}CuO_4}$&0.44&3.37&$\sim kT$ &0.9&6.78 &2.29 &8.37\\
${\rm La_{1.88}Sr_{0.12}CuO_4}$&0.57&2.78&$\sim kT$ &1.3&6.99 &3.36&4.14\\
${\rm La_{1.85}Sr_{0.15}CuO_4}$&0.7&3.11&$\sim kT$ &2.4&6.56 &2.69&1.14\\
${\rm La_{1.80}Sr_{0.20}CuO_4}$&0.87&3.01&$\sim kT$ &5.5&6.67 &2.87&0.22\\
\hline
$\bsco$&1.25&0.88& $\sim kT/2$& 0.1 & 2.64&16.8&264\\
\end{tabular}
\end{table}


\end{document}